\documentclass[twocolumn,showpacs,preprintnumbers,amsmath,amssymb]{revtex4}

\usepackage{graphicx}
\usepackage{dcolumn}
\usepackage{bm}

\begin{document}


\title{Thermoelectric properties of the bismuth telluride nanowires \\in the constant-relaxation-time approximation }

\author{Igor  Bejenari}
 \email{bejenari@iieti.asm.md}
\affiliation{
Institute of Electronic Engineering and Industrial Technologies, ASM, 3/3
Academiei str., MD2028 Kishinev, Moldova}
\affiliation{Bogoliubov Laboratory of Theoretical Physics, \\Joint Institute for Nuclear Research, 141980 Dubna, Moscow region, Russia}
 \homepage{http://theor.jinr.ru/disorder/bejenari.html}

\author{Valeriu Kantser}%
 \email{kantser@iieti.asm.md}
\affiliation{
Institute of Electronic Engineering and Industrial Technologies, ASM, 3/3
Academiei str., MD2028 Kishinev, Moldova}

\date{\today}

\begin{abstract}
Electronic structure of bismuth telluride nanowires with the growth directions [110] and [015] is studied in the framework of the anisotropic effective mass method using the parabolic band approximation. The components of the electron and hole effective mass tensors for six valleys are calculated for both growth directions. For a square nanowire, in the temperature range from 77 K to 500 K, the dependence of the Seebeck coefficient ${S}$, the thermal  ${\kappa}$ and electrical  conductivity ${\sigma}$ as well as the figure of merit ${ZT}$ on the nanowire thickness and on the excess hole concentration ${p_{ex}}$ are investigated in the constant-relaxation-time approximation. The carrier confinement is shown to play essential role for nanowires with cross section less than ${30\times 30\:\text{nm}^2}$. In contrast to the excess holes (impurities), the confinement decreases both the carrier concentration and the thermal conductivity but increases the maximum value of the Seebeck coefficient. The confinement effect is stronger for the direction [015] than for the direction [110] due to the carrier mass difference for these directions. In the restricted temperature range, the size quantum limit is valid when the ${\bm p \--}$type nanowire cross section is smaller than ${8\times 10\:\text{nm}^2}$ (${6\times 7\:\text{nm}^2}$ and ${5\times 5\:\text{nm}^2}$) at the excess hole concentration ${p_{ex}=2\times 10^{18}\text{cm}^{-3}}$ (${p_{ex}=5\times 10^{18}\text{cm}^{-3}}$ and ${p_{ex}=1\times 10^{19}\text{cm}^{-3}}$ correspondingly). The carrier confinement increases the maximum value of ${ZT}$ and shifts it towards high temperatures. For the growth direction [110], the maximum value of the figure of merit for the ${\bm p\--}$type nanowire is equal to 1.4, 1.6, and 2.8, correspondingly, at temperatures 310 K, 390 K, and 480 K and the cross sections ${30\times 30\: \text{nm}^2}$, ${15\times 15\: \text{nm}^2}$, and ${7\times 7\: \text{nm}^2}$ (${p_{ex}=5\times10^{18}\: \text{cm}^{-3}}$). At the room temperature, the figure of merit equals 1.2, 1.3, and 1.7, respectively.\end{abstract}

\maketitle

\section{\label{sec:level1}Introduction}

The study of nanowire systems is of interest because of their possible applications, in particular, in high-efficiency thermoelectric devices. The strong two-dimensional confinement of such systems allows manipulating kinetic effects \cite{Dresselhaus_FTT41,Hicks_PRB47_24,Zhou_APL87}. The reduction of dimensions leads: (i) to an increase in the Seebeck coefficient because of the increasing density of states in the vicinity of the Fermi energy, (ii) to a possible use of the anisotropy factor of the Fermi surface in the multi-valley semiconductors, (iii) to an increase in the phonon boundary scattering at interfaces of a heterostructure without an appreciable increase in electron boundary scattering, (iv) to an increase in the carrier mobility at a given concentration due to the size quantization effect. The most promising thermoelements for cooler manufacturing are those based on the nanostructures consisting of anisotropic materials like bismuth and lead tellurides  \cite{Heremans_PRL88,Heremans_PRB70}. Bismuth telluride is such a multi-valley material with a highly anisotropic isoenergy surface near the ${L\--}$point in the Brillouin zone. The charge carrier effective mass anisotropy enhances the carrier confinement effect and complicates the carrier motion through a cross section of the nanowire. The anisotropy factor leads to a modification of physical properties of the nanowire. For instance, the semimetal-semiconductor transition takes place at the Bi nanowire diameter dependent on the effective mass anisotropy \cite{Bejenari_SST19,Lin_PRB62}. Bismuth telluride and its solid solutions (${Bi_{2-x}Sb_x Te_3}$, ${Bi_2Te_{3-y}Se_y}$) are presently the best thermoelectric materials for commercial applications at room temperature. A large value of the thermoelectric efficiency ${Z}$ of these materials is due to the high degeneracy of the energy band edges. It is necessary to increase the Seebeck coefficient or the electric conductivity in order to increase the thermoelectric efficiency. An increase in the Seebeck coefficient is provided by the increase in the carrier effective mass, while the electric conductivity increases due to a decrease in the effective mass. The effective mass anisotropy factor allows for bypassing this problem and satisfying both conditions. Fast development of nanotechnologies provided various techniques to prepare bismuth telluride nanowires. Monocrystal and polycrystal ${Bi_x Te_{1-x}}$ nanowires with diameters ${40 \div 60 \:\text {nm}}$ are obtained by means of the electrochemical deposition in the nanopores of anodized alumina membranes \cite{Zhou_APL87,Li_Nanotech17}. Freestanding Bi and ${Bi_2 Te_3}$ nanowires are fabricated using both the Ulitovsky technique and the method of a high-pressure injection of the melt into capillaries \cite{Leporda_MJPS3}. The diameter of such nanowires varies from 100 nm to ${10 \:\mu \text m}$. There are only two growth directions [110] and [015] for bismuth telluride nanowires. 

The thermal conductivity ${\kappa}$ of ${Bi_2 Te_3}$ nanowires with diameter 40 nm was experimentally shown to be reduced at least by an order of magnitude from the bulk value due to the phonon-boundary scattering, which dominates phonon-phonon Umklapp scattering at room temperature \cite{Zhou_APL87}. The Seebeck coefficient ${S}$ was measured to be by ${15\%\div 60\%}$ larger than the bulk values \cite{Borca_APL85}. This thermal conductivity reduction and the aforementioned increase in the Seebeck coefficient in ${Bi_2 Te_3}$ nanowires can be exploited for enhancing the thermoelectric figure of merit. There are only a few experimental results related to thermoelectric parameters of bismuth telluride nanowires.

Here, we consider thermoelectric parameters of the rectangular bismuth telluride nanowires with growth directions [015] and [110] using a constant-relaxation-time approximation, when both carrier energy and momentum are conserved in scattering processes. The carrier mobility coincides with the bulk value. The calculation method of kinetic coefficients in this case is similar to that treated by Lin for monopolar cylindrical Bi nanowires \cite{Lin_PRB62}. We extended this method for an intrinsic semiconductor nanowire, taking into consideration the most of energy subbands in the bulk conduction and valence bands. The relaxation time depends on the carrier energy beyond the constant-relaxation-time approximation as  ${\tau(\varepsilon)=\tau_0 \varepsilon^r}$. Fitting the experimental data to the theoretical expression for the Seebeck coefficient gives the scattering factor {\it r} which varies from  -1/2 to 1/2 in the bulk bismuth telluride material when ${77\: \text K \textless T \textless \:300 \:\text K}$ \cite{Kulbachinskii_PSS199}.  Therefore, our approximation, ${r=0}$, is adequate for bismuth telluride nanowires when ${T\textgreater 77\: \text K}$. 
 
Quantum well structures and nanowires based on ${Bi_2 Te_3}$ have been studied previously in the size quantum limit (SQL) taking into account only the lowest subband using the bulk-effective-mass approximation \cite{Hicks_PRB47_24,Hicks_PRB47_19}. The nanowire growth direction was supposed to be along one of the crystallographic axes. We recall that the efficiency of a thermoelectric material is characterized by means of the figure of merit ${ZT=\sigma TS^2/\kappa}$, where ${\sigma}$ is the electrical conductivity. The maximum value of the figure of merit was obtained to be equal to ${ZT=5}$ for a ${5\AA}$-thick quantum well and ${ZT=14}$ for a ${5\AA}$-wide quantum wire. ${ZT}$ significantly increases when the square nanowire width becomes smaller, by the order of magnitude, than the thermal de Broglie wavelength  ${\lambda_D \approx(h^2/2mk_B T)^{1/2}}$. We will show that the SQL is valid only for nanowires with very small cross section (${8\times 8\:\text{nm}^2}$) in a restricted temperature range. To treat the nanowires with larger cross sections at temperature 77 K and higher, we consider all subbands of the conduction and valence bands. For the calculation, we mainly use conventional bulk ${Bi_2 Te_3}$ parameters with the exception of the lattice thermal conductivity, which is less by an order of magnitude than the bulk value. Nevertheless, we suppose the temperature dependence of the phonon thermal conductivity to be similar to that for the bulk material because of the absence of corresponding experimental data. The carrier confinement leads to a splitting of electronic band structure into subbands, while the phonon dispersion is assumed to be unchanged. The boundary scattering is not taken into account in our calculations. In general, we have improved the values of thermoelectric parameters of ${Bi_2 Te_3}$ nanowires obtained with the constant-relaxation-time approximation and provided a background for more sophisticated calculations with account of different scattering mechanisms.

In the next section, we consider the longitudinal and cross-sectional effective mass components as well as the electronic band structure of six carrier pockets in the parabolic-band approximation for the rectangular bismuth telluride nanowires with growth directions [015] and [110]. The parabolic-band approximation is adequate for a calculation of the electron energy spectrum because bismuth telluride is an indirect semiconductor. We take into account the temperature dependence of the effective mass of the carriers, which is appropriate for the intrinsic bismuth telluride. The temperature dependence of the electron and hole subbands is also considered to obtain more realistic values of the temperature dependent thermoelectric parameters. We analyze conditions at which the carrier confinement effect on transport properties of bismuth telluride nanowires can be experimentally fixed. We also consider the nanowire cross section and the excess hole concentration at which the size quantum limit is adequate. In Sec.3, a dependence of the Fermi energy and the carrier concentration on temperature is studied for the nanowires with different cross sections and the excess hole concentrations. In Sec. 4, the temperature dependence of the Seebeck coefficient, the thermal conductivity, and the figure of merit are studied. Finally, conclusions are given in Sec. 5. 
 
\section{\label{sec:level1} Electronic Structure}
\subsection{\label{sec:level2} Electronic Band Structure of the Bulk Material}

The crystal structure of bismuth telluride with rhombohedral unit cell belongs to the symmetry group ${D^5_{3d}}$ ${(R\bar{3}m)}$ \cite{Goltsman,Goldsmid}. The crystallographic axes are binary ${\bm{n}\:(x)}$, bisectrix ${\bm{s}\:(y)}$, and trigonal ${\bm{c}\:(z)}$. The electronic band structure of the indirect band gap ${Bi_2 Te_3}$ semiconductor is given by a parabolic-band approximation in the Drable-Wolf six-valley model. All six valleys are equivalent in the bulk. The tilt angle between the hole (electron) energy ellipsoid principle axes and bisectrix-binary plane is ${32^\circ}$ (${34^\circ}$) \cite{Kohler_PSS74,Kohler_PSS73}.The measurement of the Shubnikov-de Haas effect showed the bulk effective mass tensor components to be ${m_{h1}=0.0308 \:m_0}$, ${m_{h2}=0.442 \: m_0}$, and ${m_{h3}=0.0862\: m_0}$  for holes and ${m_{e1}=0.0213\: m_0}$, ${m_{e2}=0.319\: m_0}$, and ${m_{e3}=0.0813\: m_0}$ for electrons in the local coordinate system of the ellipsoid. The energy band gap was obtained by means of the measurement of the absorption coefficient and the electrical conductivity. The indirect (direct) band gap is equal to 0.15 eV (0.22 eV) at temperature 2 K. It is 0.13 eV at room temperature, being about linearly dependent on temperature with the rate ${dE_g/dT=-0.09}$ meV/T. This agrees rather well with the recent calculations made within the screened-exchange local density approximation \cite{Kim_PRB72}. Using the general formula ${E_g(T)=E_g(0)-\alpha T^2/(T+\beta)}$ reliable for most semiconductors and the experimental data listed above we obtained the following expression for temperature dependence of ${Bi_2 Te_3}$ energy band gap \cite{Nag}
\begin{equation}
E_g(T)=150-\frac{0.0947\:T^2}{T+122.5}\:[\text{meV}]
\label{eq:1}.
\end{equation}

The electrical conductivity measured in the bismuth telluride bulk material varies with temperature as  ${\sigma(T)\approx T^{-2.0}\div T^{-1.2}}$ when the carrier concentration is about ${10^{19}\text{cm}^{-3}}$ \cite{Drabble_PPS72_401, Drabble_PPS72_380, Goldsmid_PPS71}. For classical statistics, when the acoustic phonon scattering is dominant, the electrical conductivity depends on temperature as  ${\sigma(T)\approx m^{-5/2}T^{-3/2}}$ \cite{Kutasov_FTT39}. Hence, in the case of the intrinsic conductivity, the temperature dependence of the effective mass is given by ${m_h\approx T^{0.14} \div T^{0.2}}$  for holes and ${m_e\approx T^{0.12}}$ for electrons \cite{Kutasov_FTT7}. 

The parabolic valence band structure is described by the Schrodinger equation in the rhombohedral coordinate system for the bulk bismuth telluride as follows \cite{Goltsman}
\begin{equation}
\frac{\hbar^2}{2}\mathbf{\nabla}{\mathbf {\hat{\alpha}}} \mathbf{\nabla}\Psi({\mathbf r})=E\Psi({\mathbf r})
\label{eq:2},
\end{equation}
where ${\bm{\hat{\alpha}}}$ is the inverse tensor of the hole effective mass with components   ${\alpha_{11}=32.5\:m^{-1}_0}$, ${\alpha_{22}=4.81\:m^{-1}_0}$, ${\alpha_{33}=9.02\:m^{-1}_0}$, and ${\alpha_{23}=4.15\:m^{-1}_0}$ \cite{Kohler_PSS74}. The edge of the ${Bi_2 Te_3}$ valence band is displaced from the edge of the conduction band by the vector ${\mathbf {k_0}=(0.091, 0.152, 0.152)\: \text{nm}^{-1}}$ in the momentum space \cite{Kim_PRB72}. Due to this displacement, the conduction band structure is described as 
\begin{equation}
\frac{\hbar^2}{2}(-i\mathbf{\nabla}-\mathbf{k_0}){\mathbf {\hat{\alpha}}} (-i\mathbf{\nabla}-\mathbf{k_0})\Psi({\mathbf r})=E\Psi({\mathbf r})
\label{eq:3}.
\end{equation}
The components of the inverse electron effective mass tensor are  ${\alpha_{11}=46.9\:m^{-1}_0}$, ${\alpha_{22}=5.92\:m^{-1}_0}$, ${\alpha_{33}=9.50\:m^{-1}_0}$, and ${\alpha_{23}=4.22\:m^{-1}_0}$ \cite{Kohler_PSS73}. Scattering of charge carriers from different valleys (${X\rightarrow L}$) is similar to that from equivalent valleys, the difference of the carrier effective masses and the edges of different valleys being taken into account  \cite{Ridley}. Hence, the displacement vector ${\mathbf{k_0}}$ of the edges can be neglected in the calculation of kinetic coefficients.  

\subsection{\label{sec:level3} Anisotropic Band Structure in the Bismuth Telluride  Nanowire}

The growth direction [110] of a ${Bi_2 Te_3}$ nanowire with diameter ${50 \div 100 \:\text {nm}}$ is perpendicular to the trigonal axis. This direction is equivalent to ${[11\bar{2}0]}$ in the hexagonal unit cell.  It is parallel to one of three bisectrix axes perpendicular to the trigonal axis \cite{Zhou_APL87}. We choose the nanowire coordinate system with $y$ axis along the bisectrix axis to apply the boundary condition to Eqs.~(\ref{eq:2}) and (\ref{eq:3}). We suppose that the wave function vanishes at the nanowire boundary. Two direct equivalent valleys 1 and 4 are distinguished from the set of four oblique equivalent valleys 2, 3, 5, and 6 for the [110] growth direction. Since the carrier motion in the $y$-direction is free, ${\Psi_{1,4}(\mathbf r)=u(x,z)\exp [{}i k_y(y-z\alpha^2_{23}/\alpha_{33})]}$  is an appropriate wave function of the hole pockets 1 and 4 aligned with the bisectrix axis and Eq.~(\ref{eq:2}) can be rewritten as
\begin{equation}
\frac{\hbar^2}{2}\left(\alpha_{11}\frac{\partial^2u}{\partial x^2}+\alpha_{33}\frac{\partial^2u}{\partial z^2}\right)=\left(E+\frac{\hbar^2 k^2_y}{2m^*_y}\right)u(x,z)
\label{eq:4},
\end{equation}
where the longitudinal effective mass component is ${m^*_y=(\alpha_{22}-\alpha^2_{23}/\alpha_{33})^{-1}}$. The transverse effective mass components are ${m_x=1/\alpha_{11}}$ and ${m_z=1/\alpha_{33}}$.

For a self-consistent calculation of the energy spectra for different valleys, we use a  (rhombohedral) coordinate system associated with the nanowire. Therefore, one has to rotate the hole (electron) pockets 2, 3, 5, and 6 about ${z\:(\mathbf c)}$ axis to obtain the corresponding inverse effective mass tensor ${\hat{\alpha}^{(p)}={\left[\hat R^{(p)}_{xy}\right]}^{-1} \hat{\alpha} \hat R^{(p)}_{xy}}$  in the rhombohedral system. The rotation operator is 
\begin{equation}
\hat R^{(p)}_{xy}=\left(
\begin{array}{ccc} %
								\cos[(p-1)\pi/3]  & \sin[(p-1)\pi/3] &  0 \cr
								-\sin[(p-1)\pi/3] &  \cos[(p-1)\pi/3]	 &	0	\cr
								0									&				0					 &  \lambda
\end{array}\right)
\label{eq:5}.
\end{equation}
The parameter $p$ indicates the number of the corresponding valley (1, 2,\ldots, 6);  ${\lambda=+1}$ if ${p=1, 2, 3}$ and  ${\lambda=-1}$ if ${p=4, 5, 6}$. To eliminate the mixed partial derivative from the Schrodinger equation for the oblique valleys, one has also to transform the rhombohedral coordinate system into a new one $(x', y', z')$ by using the rotation operator  ${\hat R_{xz}}$ about ${y}$ (or ${\mathbf s}$) axis by the angle ${\Theta}$ defined from the relation ${\tan (2\Theta)=\pm 2\alpha^{(p)}_{13}/(\alpha^{(p)}_{33}-\alpha^{(p)}_{11})}$. In the new coordinate system, the wave function for the oblique valleys is
\begin{eqnarray}
F^{(p)}(\mathbf {r'})=&&u(x',z')\exp(ik_yy)\exp\left(-i\frac{\beta^{(p)}_{12}k_y}{\beta^{(p)}_{11}}x'\right)
 \nonumber \\
&&\times \exp\left(-i\frac{\beta^{(p)}_{23}k_y}{\beta^{(p)}_{33}}z'\right)
\label{eq:6}.
\end{eqnarray}
A function ${u(x',z')}$ satisfies the equation in form (\ref{eq:4}) with a transformed inverse effective mass tensor ${\hat \beta^{(p)}=\hat R^{-1}_{xz}\hat \alpha^{(p)} \hat R_{xz}}$. Therefore, the longitudinal effective mass component for the oblique valleys 2, 3, 5, and 6 is
\begin{equation}
\frac{1}{m^{(p)*}_y}=\beta^{(p)}_{22}-\frac{\left[\beta^{(p)}_{12}\right]^2}{\beta^{(p)}_{11}}
-\frac{\left[\beta^{(p)}_{23}\right]^2}{\beta^{(p)}_{33}}
\label{eq:7}.
\end{equation}
The transverse effective mass components are ${m^{(p)}_x=1/\beta^{(p)}_{11}}$  and ${m^{(p)}_z=1/\beta^{(p)}_{33}}$ .

\begin{table*}
\caption{\label{tab:1} Calculated effective mass components of a \emph{hole} pocket in the bismuth telluride nanowire along the indicated crystallographic directions. The $y$ direction is chosen along the nanowire growth direction (the wire axis). All values of the mass components are in units of the free electron mass ${m_0}$.}
\begin{ruledtabular}
\begin{tabular}{ccccccc}
 &\multicolumn{3}{c}{Bisectrix}&\multicolumn{3}{c}{${[015]}$}\\
Hole mass&1st and 4th&2nd and 5th&3rd and 6th
&1st and 4th&2nd and 5th&3rd and 6th\\
component&pocket&pocket&pocket&pocket&pocket&pocket\\ \hline
 ${m_x}$ & 0.03077 & 0.07034 & 0.07034& 0.03077 & 0.1017 & 0.4220 \\
 ${m^*_y}$ & 0.3448 & 0.1093 & 0.1093 & 0.4421 & 0.2013 & 0.05893 \\
 ${m_z}$ & 0.1109 & 0.1530 & 0.1530 & 0.08645 & 0.05744 & 0.04729 \\
\end{tabular}
\end{ruledtabular}
\end{table*}

\begin{table*}
\caption{\label{tab:2}  Calculated effective mass components of an \emph{electron} pocket in the bismuth telluride nanowire along the indicated crystallographic directions. The $y$ direction is chosen along the nanowire growth direction (the wire axis). All values of the mass components are in units of the free electron mass ${m_0}$.}
\begin{ruledtabular}
\begin{tabular}{ccccccc}
 &\multicolumn{3}{c}{Bisectrix}&\multicolumn{3}{c}{${[015]}$}\\
Electron mass&1st and 4th&2nd and 5th&3rd and 6th
&1st and 4th&2nd and 5th&3rd and 6th\\
component&pocket&pocket&pocket&pocket&pocket&pocket\\ \hline
 ${m_x}$   & 0.02132& 0.05625 & 0.05625 & 0.02132 & 0.08977 & 0.03550 \\
 ${m^*_y}$ & 0.2472 & 0.07779 & 0.07779 & 0.3197  & 0.1484  & 0.04982 \\
 ${m_z}$   & 0.1053 & 0.1268  & 0.1268  & 0.08139 & 0.04166 & 0.3137 \\
\end{tabular}
\end{ruledtabular}
\end{table*}

${Bi_{0.46}Te_{0.54}}$ nanowires with diameter ${40 \div 60}$ nm mainly grow along the [015] direction \cite{Zhou_APL87}. This direction lies along the vector ${\mathbf l=(0,0.8489,0.5285)}$ in the rhombohedral coordinate system. In general, the growth direction of the bismuth telluride nanowire tends towards the trigonal axis with decreasing diameter in the same way as a Bi nanowire does \cite{Lin_PRB62}. To satisfy the boundary conditions, Eqs.~(\ref{eq:2}) and (\ref{eq:3}) should be written in a nanowire coordinate system $(x", y", z")$ with the ${y"}$ axis along the vector ${\mathbf l}$. We have to use the rotation matrix ${\hat R_{yz}}$ to transform the rhombohedral coordinate system into the nanowire coordinate system. The rotation is about the binary axis by the angle  ${\Theta'=\arccos(0.8489)}$ ${(32^\circ)}$. Hence, this matrix has the form
\begin{equation}
\hat R_{yz}=\left(
\begin{array}{ccc} %
								1 &  0       &  0 \cr
								0 &  0.8489	 &	-0.5285	\cr
								0	&	 0.5285	 &  0.8489
\end{array}\right)
\label{eq:8}.
\end{equation}
In the new system of coordinates, the wave function and the longitudinal effective mass component for six valleys are written in the form (\ref{eq:6}) and (\ref{eq:7}), correspondingly. The components of the inverse effective mass tensor are defined as ${\hat \gamma^{(p)} = \hat R^{-1}_{xz} \hat R^{-1}_{yz} \hat \alpha^{(p)} \hat R_{yz} \hat R_{xz}}$. In this case, the rotation angle ${\Theta}$ is slightly modified taking into account the transformation ${\hat R_{yz}}$.

The transverse and longitudinal hole (electron) effective mass components in the bismuth telluride nanowires oriented along the bisectrix axis and the direction [015] are listed in Table \ref{tab:1} (\ref{tab:2}). The hole effective mass is greater than that of electron both in bulk and in the nanowire. For the nanowire growth direction [015], the carrier effective mass components for the 1st and 4th valley slightly differ from the bulk components because this direction is approximately oriented along one of the principle axis of the bulk electron (hole) ellipsoid. Since the effective mass components are alike for valleys 2, 3, 5, and 6 in the nanowire oriented along the bisectrix axis, these valleys are equivalent in this case. For the nanowires with the growth direction [015], some of the hole (electron) effective mass components in valleys 2 and 5 differ from the corresponding components in valleys 3 and 6 approximately by a factor of four (eight). The transverse effective mass anisotropy of carriers in valleys 1 and 4 (3 and 6) for the nanowires oriented along the bisectrix axis is greater (less) than that for the nanowires with the growth direction [015]. The mass anisotropy for valleys 2 and 5 is similar for both growth directions. For valleys 1, 2, 4, and 5, the cyclotron effective mass  ${\sqrt{m_x m_z}}$ for the direction [015] is less than that for nanowires oriented along the bisectrix axis. For nanowires with the growth direction [015], the effective mass anisotropy in valleys 3 and 6 is greater than that for the other direction. Due to last two factors, the confinement effect for the nanowire with the growth direction [015] is larger than that for the nanowire with the growth direction along the bisectrix axis. 

Here, we study bismuth telluride nanowires with a rectangular cross section. Similar bismuth nanowires have been recently fabricated using the electron beam lithography \cite{Chiu_Nanotech15}. The energy spectrum of holes is defined from Eq.~(\ref{eq:4}) as follows
\begin{equation}
E^{hole}_{n,l}(k_y)=-\frac{\hbar^2 \pi^2}{2}\left( \frac{n^2}{m_x a^2_x}+\frac{l^2}{m_z a^2_z} \right)-\frac{\hbar^2 k^2_y}{2m^*_y}
\label{eq:9}.
\end{equation}
The electron energy is defined from Eq.~(\ref{eq:3}) in the form
\begin{eqnarray}
E^{el}_{n,l}(k_y)=&&E_g+\frac{\hbar^2}{2}\left[ \frac{{(n\pi/a_x-k_{0x})}^2}{m_x }+\frac{{(l\pi/a_z-k_{0z})}^2}{m_z} \right] \nonumber \\
&&+\frac{\hbar^2 {(k_y-k_{0y})}^2}{2m^*_y}
\label{eq:10}.
\end{eqnarray}
Symbols ${a_x}$ and ${a_z}$ denote the sides of the rectangular cross section. The carrier effective mass components ${m_x}$, ${m^*_y}$, and ${m_z}$ are listed in Tables~\ref{tab:1} and \ref{tab:2}. The electron and hole wave functions differ by the oscillating factor ${\exp(i\mathbf{k_0r})}$ which does not effect the transport matrix elements.
\begin{figure}[h]
\includegraphics{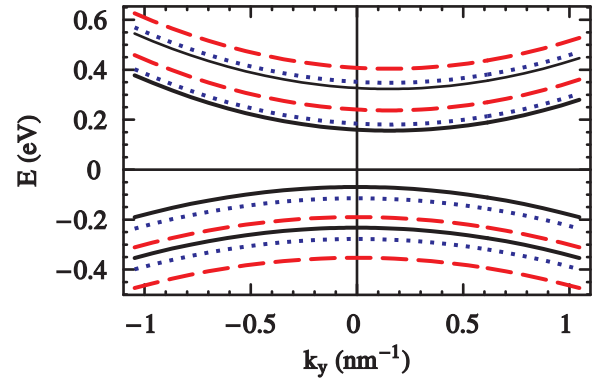}
\caption{\label{fig:1} Electron and hole energy spectrum for valleys 1 and 4 in the intrinsic nanowire ${(a_x=a_z=15 \text{nm})}$ with growth direction [110]. The solid lines correspond to quantum numbers ${n=1,2}$ and ${l=1}$, dotted lines to ${l=2}$, dashed lines to ${l=3}$.}
\end{figure}

Fig.~\ref{fig:1} shows the electron and hole energy spectrum in the intrinsic nanowire with the growth direction along the bisectrix axis at room temperature. The spectrum is composed of six subbands for equivalent valleys 1 and 4. The sides of the nanowire cross section are ${a_x=a_z=15}$ nm. The energy origin is taken at the top of the valence band in the bulk. The bottom of the electron subbands is displaced from the top of the valence subbands along the axis ${k_y}$ by 0.152 nm. Fig.~1 demonstrates that the size quantization is manifested more significantly for holes than for electrons. However, such a large difference between the electron and hole confinement is an apparent effect. The real difference is smaller. The sketch in Fig.~1 is reflected when the energy origin is situated at the bottom of the bulk conduction band, while the confinement effect does not depend on the choice of the energy origin. A slight difference between the electron and hole confinement exists because the electron and hole masses are not equal. The energy separation between the subbands with different quantum numbers  ${n}$ is greater than that between the subbands with different quantum numbers ${l}$ due to the ratio of the transversal mass components ${m_x}$ and ${m_z}$. The numerical calculations demonstrate that a proper choice of the sides of the rectangular cross section allows for manipulating the energy subband splitting. A fitting of the ratios of the sides diminishes or enhances the mass anisotropy effect on the energy spectrum. This agrees with the conclusion of our previous research on Bi nanowires \cite{Bejenari_SST19}. We have thus obtained that the model of carriers with an anisotropic mass in the nanowire with a circular cross section is identical with the model of carriers with an isotropic mass in the nanowire with an elliptic cross section. 

\subsection{\label{sec:level4}  Evaluation Criteria of a Quantum Size Effect in the Bismuth Telluride Nanowire}

An experimental observation of the quantum size effect is possible at some favorable conditions. These conditions are mainly based on three criteria. The first criterion states that a thermal excitation should be small enough to provide interband carrier transitions, which diminish the effect of the band splitting on the transport properties. In the so-called "classical limit", if the thermal energy ${k_BT}$ is much larger than the energy level spacing  ${\Delta E}$, the subbands can be treated as continua owing to the thermal smearing of states \cite{Lin_PRB68}. In the "quantum regime", variations in the current peak height versus gate voltage measured, for example, in Si nanowire device suggest that the transport is carried out through coherent energy states with the energy level spacing larger than the thermal energy \cite{Lu_JPDAP39}. The subband splitting increases with increase of quantum numbers in the model of a cylindrical potential well with a rectangular cross section. Hence, taking into account Eq.~(\ref{eq:9}), the first criterion can be mathematically expressed as
\begin {equation}
max[E_{2,1}(k_y),E_{1,2}(k_y)]-E_{1,1}(k_y)\gg k_B T
\label{eq:11}.
\end{equation}
We choose the maximum energy values of subbands with quantum numbers (1,2) and (2,1) in the relation~(\ref{eq:11}) because the transition matrix elements including the respective wave functions ${F_{12}(\mathbf r)}$ and ${F_{21}(\mathbf r)}$ are slightly different. In the case of a degenerate semiconductor, the quantum size effect is reduced when there are many electron subbands under the Fermi level. Since the carrier thermal excitation is much less than the Fermi energy, the first criterion for a degenerate nanowire reads
\begin {equation}
E_{n,l+1}(k_y)-E_{n,l}(k_y)\geq E_F
\label{eq:12},
\end{equation}
where the quantum numbers $n$ and $l$ are small. Conditions (\ref{eq:11}) and (\ref{eq:12}) are equivalent to the requirement that the de Broglie wavelength (${\lambda_D}$) of electrons should be of order of or larger than the nanowire thickness $a$ \cite{Lin_PRB68}
\begin {equation}
\frac{\lambda_D}{a}\geq1
\label{eq:13}.
\end{equation}
In the case of classical statistics, the de Broglie wavelength is defined as ${\lambda_D\approx {2\pi\hbar}/\sqrt{2mk_BT}}$ and the condition (\ref{eq:11}) can be obtained from Eq.~(\ref{eq:13}). Taking into account that ${\lambda_D\approx\lambda_F=2\pi\hbar/\sqrt{2mE_F}}$  for a degenerate semiconductor nanowire, the condition~(\ref{eq:13}) can be transformed into the expression~(\ref{eq:12}).
\begin{figure}[h]
\includegraphics{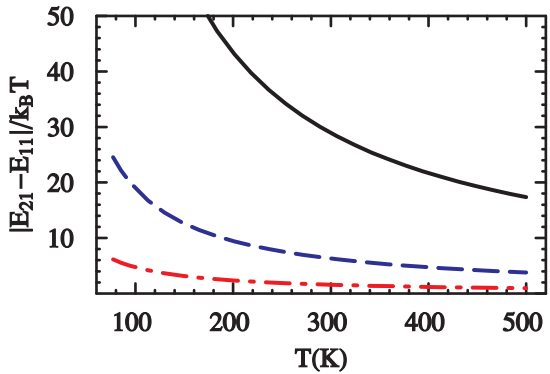}
\caption{\label{fig:2} Temperature dependence of the reduced energy separation between the hole subbands (1,1) and (2,1) in the square nanowire, with thickness ${a=30\: \text{nm}}$ (dashed-dotted line), 15 nm (dashed line), and 7 nm (solid line) for the 1st and 4th valley.}
\end{figure}

Fig.~\ref{fig:2} demonstrates the temperature dependence of the energy separation between the hole subbands with quantum numbers (1,1) and (2,1) in the units of thermal energy in the square nanowires with growth orientation [110]. The subband separation in the nanowires with thickness 30 nm tends to unity when ${T > 350}$ K. The first criterion is satisfied in the nanowires with thickness less than 30 nm for whole temperature range, ${77\: \text K<T<500 \: K}$. Hence, the carrier confinement affects the transport properties in the nanowires with cross section less than ${30\times30 \: \text{nm}^2}$. The results for both nanowire growth directions are approximately similar.

An electron (hole) can occupy the quantum state during a certain interval of time ${\tau}$ because of scattering. The final lifetime of the quantum state leads to uncertainty of the corresponding energy due to the energy-time uncertainty principle. Energies $E$ and ${E+\Delta E}$ cannot be resolved if the mean time between collisions ${\tau}$ is shorter than that given by the energy-time uncertainty ${\Delta E\tau \geq\hbar}$  because of a collision broadening of the energy levels by  ${\Delta E}$ \cite{Scholl}. Therefore, the second criterion is that the space between the discrete energy levels must be much greater than the energy uncertainty, in order to keep the peculiarity of the one-dimensional density of states in nanowire. It is mathematically expressed for a rectangular nanowire as
\begin{equation}
E_{2,1}-E_{1,1}\gg \frac{\hbar}{\tau}\approx\frac{\hbar e}{m \mu}
\label{eq:14}.
\end{equation}
To estimate the criterion, we use the value of the electron (hole) mobility  ${\mu^{[110]}_e=1200\:\text{cm}^2/\text{Vs}}$  (${\mu^{[110]}_h=510\:\text{cm}^2/\text{Vs}}$) obtained from the measurement of the electrical conductivity in the cleavage plane of a ${Bi_2Te_3}$ monocrystal at 300 K \cite{Goltsman, Goldsmid, Goldsmid_2}. The temperature dependence of the electron (hole) mobility is given by the relation  ${\mu_e\approx T^{-1.7}}$ (${\mu_h\approx T^{-2.0}}$) \cite{Kulbachinskii_PSS199, Fleurial_JPCS49, Champness_CJP44}. Using the mass components listed above, we estimate the electron and hole density-of-state effective mass  ${m_{DOS}=N^{2/3}_{deg}(m_1m_2m_3)^{1/3}}$ for the six-fold degenerate (${N_{deg}=6}$) bulk valleys to be equal to 0.271 ${m_0}$ and 0.348 ${m_0}$, respectively. Hence, the electron (hole) energy uncertainty ${\hbar/\tau}$ varies from 6.5 meV (3.6 meV) to 16 meV (10 meV) in the temperature range ${300\: \text K <T<500\: \text K}$. The temperature dependence of the effective mass leads to a shrinkage of limits of the energy uncertainty interval. The electron (hole) thermal excitation ranges from 26 meV to 43 meV in the same temperature interval. Therefore, the second criterion~(\ref{eq:14}) is always satisfied to observe experimentally the confinement effect in the bismuth telluride nanowires, when the first criterion~(\ref{eq:11}) or (\ref{eq:12}) is fulfilled.

In the ballistic region, the electron scattering at the surface roughness of a quantum wire destroys the conductance quantization \cite{Sone_SST7}. The third criterion requires the magnitude of roughness of the nanowire boundary to be less than the de Broglie wavelength of electrons, in order to diminish the role of the surface diffusion \cite{Ziman, Greiner_SST13}. The wavelength lower limit is ${\lambda_D=21.4\:\text{nm}}$ (13.6 nm) for the electrons and ${\lambda_D=19.5\:\text{nm}}$ (11.7 nm) for the holes at 300 K when the effective mass components are 0.1268 ${m_0}$ (0.3137 ${m_0}$) and 0.153 ${m_0}$ (0.422 ${m_0}$), respectively, for the growth direction [110] ([015]). This result agrees well with the recently estimated thermal de Broglie wavelength of holes (${\lambda_D=11.4\:\text{nm}}$) in the ${Bi_2Te_3/Sb_2Te_3}$ superlattices \cite{R_Nature}. Since the surface roughness of ${Bi_2Te_3}$  nanowires (films) is of the order of 1 nm (10 nm), the third criterion is satisfied \cite{Zhou_APL87,Zimmer}. 

The three criteria considered above include both the carrier effective mass and the nanowire cross section area. To study the size quantum limit, we also have to take into account the excess hole concentration. SQL is achieved when all electrons (holes) occupy the lowest (highest) subband in a degenerate semiconductor nanowire. This condition is mathematically expressed, starting from the relation~(\ref{eq:12}), as
\begin{equation}
n_{1D}<\frac{\sqrt{8m^*_y(E_{21}-E_{11})}}{\pi\hbar}
\label{eq:15},
\end{equation}
where the right hand side represents a full number of states in the 1st subband per unit length, ${G_{1D}(E_{21})}$. For the ${\bm p\--}$type degenerate bismuth telluride nanowires, the hole concentration ${p_{1D}}$ coincides with the excess hole concentration ${p_{ex}}$. Using the relation~(\ref{eq:15}), we obtain the maximal nanowire cross section thickness ${a_x}$ and width ${a_z}$ at which the SQL takes place, from the following relation
\begin{equation}
a^{max}_{x,z}=\left(\frac{12m^*_y}{m_{x,z}p^2_{1D}}\right)^{1/6}
\label{eq:16}.
\end{equation}
For nanowires with the growth direction [110], the maximum thickness and width is 8 nm (7 nm) and 10 nm (8 nm) at a low temperature when the excess hole concentration is ${5\times10^{18}\: \text{cm}^{-3}}$ (${1\times10^{19}\: \text{cm}^{-3}}$). For the nanowire growth direction [015], we estimate the maximum thickness and width as 6 nm (5 nm) and 9 nm (7 nm), correspondingly. Hence, the SQL for nanowires grown along [110] with a small excess hole concentration is larger than that for the growth direction [015] because the longitudinal effective mass component of holes ${m^*_y}$ in the highest valence subband for the direction [110] is twice as large as that for the other direction (for valleys 3 and 6). At a high temperature, the maximal cross section sides depend on temperature when the hole concentration deviates from the excess hole concentration. In the next sections, we study thermoelectric properties of nanowires with the cross sections ${7\times 7\: \text{nm}^2}$ and ${15\times 15\: \text{nm}^2}$ to consider both the size quantization effect and the size quantum limit for temperatures ranging from 77 to 500 K.

\section{\label{sec:level5} Carrier Statistics in the Nanowire}

\begin{figure*}
\includegraphics{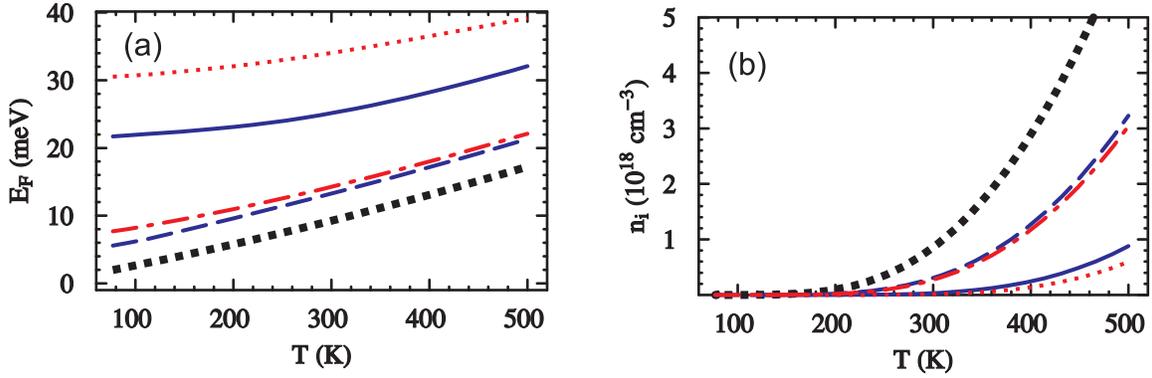}
\caption{\label{fig:3} The temperature dependence of (a) the Fermi energy and (b) the carrier concentration in the square intrinsic ${Bi_{0.37}Te_{0.63}}$ nanowire with thickness 7 nm (solid line), 15 nm (dashed line) and the growth direction [110]. Dashed-dotted and dotted lines correspond to thickness 15 nm and 7 nm for the growth direction [015]. Filled squares show the Fermi energy and the carrier concentration in the bulk.}
\end{figure*}  
Carrier concentration for an electron pocket in the nanowire is defined as \cite{Hanson, Lundstrom}
\begin{equation}
n_{1D}(E_F)=N^{1D}_{c,v}\sum\limits_{n=1}^{n_{max}}\sum\limits_{l=1}^{l_{max}(n)}\Phi_{-1/2}(\eta_{n,l})
\label{eq:17}.
\end{equation}
The factor ${N^{1D}_{c,v}=\left(2m^{e,h}_y k_B T/\pi \hbar^2\right)^{1/2}}$  denotes the effective density of states for the corresponding electron (hole) pocket in the nanowire. 
\begin{equation}
\Phi_j(\eta)=\frac{1}{\Gamma(j+1)}\int\limits_0^\infty \frac{\epsilon^j}{\exp(\epsilon-\eta)+1}d\epsilon
\label{eq:18}
\end{equation}
is the complete Fermi-Dirac integral with a fractional index. The reduced chemical potentials  $\eta$ for an electron subband and a hole subband are defined as
\begin{eqnarray}
\eta^c_{n,l}=\frac{E_F-E_c-E^e_{n,l}}{k_B T},\\
\eta^v_{n,l}=\frac{E_v+E^h_{n,l}-E_F}{k_B T}
\label{eq:19},
\end{eqnarray}
where ${E^{e(h)}_{n,l}}$ determines the electron (hole) subband edge in the nanowire relative to the bottom (top) of the conduction (valence) band for  the bismuth telluride bulk material.

The Fermi energy ${E_F}$ is calculated from the equation of the electrical neutrality ${n(E_F)=p(E_F)-p_{ex}}$, where ${p_{ex}}$ denotes the concentration of uncompensated acceptors (excess holes). The excess holes are absent in an intrinsic nanowire. The calculation of the bismuth telluride conduction (valence) band structure in the screened-exchange local density approximation demonstrated that its width ${E_{bw}}$ varies in the range ${0.8\div 0.9}$ eV \cite{Kim_PRB72}. We suppose that all electron (hole) subbands in the nanowire are fitted into the conduction (valence) band with width corresponding to the bulk material. In fact, we can take ${E_{bw}=0.6}$ eV because it is greater than ${10\:k_BT}$. The maximum values of the subband quantum numbers ${n_{max}}$ and ${l_{max}}$ included in Eq.~(\ref{eq:17}) are defined from the following expressions
\begin{eqnarray}
n_{max}&\approx&\left[\frac{2E_{bw}m_xa^2_x}{\hbar^2\pi^2}-\frac{m_x}{m_z}\left(\frac{a_x}{a_z}\right)^2\right]^{1/2},\\
l_{max}(n)&\approx&\left(\frac{2E_{bw}m_xa^2_x}{\hbar^2\pi^2}-n^2\right)^{1/2}\left(\frac{m_z}{m_x}\right)^{1/2}\frac{a_z}{a_x}. \nonumber\\
\label{eq:21}
\end{eqnarray}
 
The bulk ${Bi_xTe_{1-x}}$ material is of ${\bm p\--}$type for ${x>0.37}$ and of ${\bm n\--}$type for  ${x<0.37}$ \cite{Goltsman}. The dependence of the nanowire conductivity type on the Bi-to-Te ratio differs from that of the bulk  \cite{Zhou_APL87}. For the sake of simplicity, we suppose that the ${Bi_{0.37}Te_{0.63}}$ nanowire is intrinsic. This does not actually influence the results because the parameter $x$ is not taken into account in the calculation. 
\begin{figure*}
\includegraphics{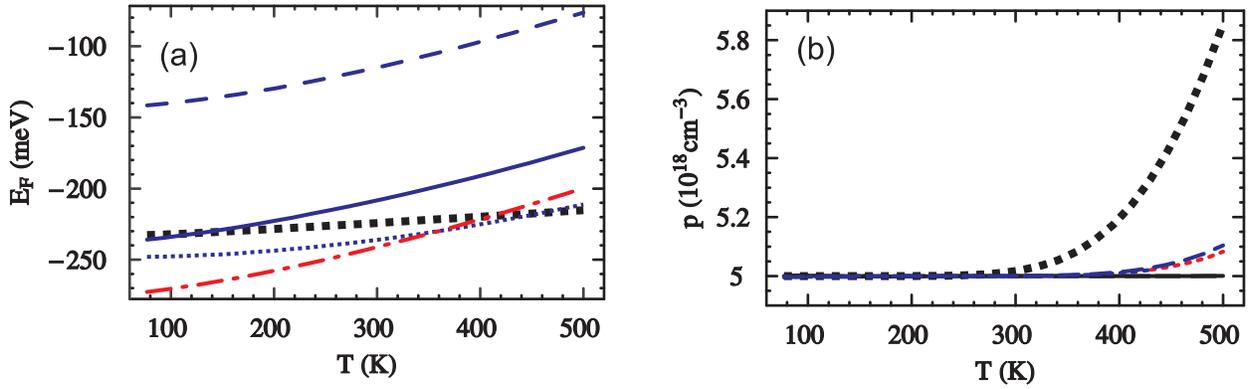}
\caption{\label{fig:4} Temperature dependence of (a) the highest hole subband (filled squares) and the Fermi energy for the square ${\bm p\--}$type ${Bi_2Te_3}$ nanowire with thickness 15 nm (dashed line) and 7 nm (solid line) at ${p_{ex}=5\times10^{18}\:\text{cm}^{-3}}$ and ${p_{ex}=1\times10^{19}\:\text{cm}^{-3}}$ (dotted line) for the growth direction [110], (b) the hole concentration in the bulk (filled squares) and in the nanowire with thickness 7 nm (solid line) and 15 nm (dashed line) for the growth direction [110] at ${p_{ex}=5\times10^{18}\:\text{cm}^{-3}}$. Dashed-dotted (dotted) line shows the Fermi energy (hole concentration) in the 7 nm (15 nm) thick nanowire with the growth direction [015] at ${p_{ex}=5\times10^{18}\:\text{cm}^{-3}}$.}
\end{figure*}

Fig.~\ref{fig:3} represents a dependence of the Fermi energy ${E_F}$ and the carrier concentration ${n_i=n_{1D}/(a_xa_z)}$ on temperature in square intrinsic ${Bi_{0.37}Te_{0.63}}$ nanowires with thickness 15 nm, 7 nm and the growth directions [110] and [015]. The energy origin is selected in the middle of the band gap. The carrier confinement increases the Fermi energy in the nanowire because the hole mass is greater than the electron mass. The increase of the Fermi energy is greater for the nanowire growth direction [015] compared with the other direction at the same nanowire thickness. The difference between the Fermi energies associated with the directions [015] and [110] increases when both the nanowire cross section and temperature decrease. For the ${Bi_{0.37}Te_{0.63}}$ nanowires, the Fermi energy lies in the band gap in the whole temperature range. Therefore, the carrier confinement does not change the conductivity type of such nanowires. Hence, we take into account the temperature dependence of the effective electron (hole) mass ${m_e=T^{0.12}}$ (${m_h=T^{0.17}}$) in our calculation of transport properties of the intrinsic bismuth telluride nanowires. This mass temperature dependence leads to an additional increase in the carrier concentration with temperature. Fig. 3(b) demonstrates that the confinement decreases the carrier concentration in the intrinsic nanowires. This agrees with the calculations for the Bi nanowires \cite{Lin_PRB62}. The confinement is greater for the direction [015] than for [110], therefore, the carrier concentration for the direction [015] is somewhat less than for the other direction at the same value of the nanowire cross section. This small difference increases when temperature increases or the nanowire cross section decreases. 

Fig.~\ref{fig:4} demonstrates the temperature dependence of the Fermi energy and the hole concentration for the ${\bm{p}\--}$type ${Bi_2Te_3}$ nanowires with the cross section ${15\times15\:\text{nm}^2}$ and ${7\times 7\:\text{nm}^2}$ for the both growth directions. For the nanowire with the cross section ${7\times 7\:\text{nm}^2}$ grown along the [110] direction, the highest hole subband of the equivalent valleys 2 and 6 is depicted by filled squares in Fig.~\ref{fig:4}(a). The energy origin is taken at the middle of the band gap. The bulk Fermi energy varies from  -100 meV to  -37 meV in the temperature interval ${77\: \text K <T<500\: \text K}$ at the excess hole concentration  ${p_{ex}=5\times10^{18}\:\text{cm}^{-3}}$. If the nanowire cross section area decreases by a factor of five, the Fermi energy decreases by 40 per cent (50 per cent) at low (high) temperature. For the square nanowire with thickness 7 nm and the excess hole concentration  ${p_{ex}=5\times10^{18}\:\text{cm}^{-3}}$, the Fermi energy corresponding to the [110] and [015] growth directions lies under the highest valence subband when ${T<180\:\text K}$ and ${T<400\:\text K}$, respectively. For the excess hole concentration ${p_{ex}=1\times10^{19}\:\text{cm}^{-3}}$, the Fermi energy is under the highest hole subband in the whole temperature range. The hole concentration in the nanowire depends neither on temperature nor on the cross section area at a high value of the excess hole concentration ${p_{ex}=1\times10^{19}\:\text{cm}^{-3}}$. For a smaller excess hole concentration ${p_{ex}=5\times10^{18}\:\text{cm}^{-3}}$ the hole concentration in the nanowire with the cross section ${15\times15\:\text{nm}^2}$ slightly increases with temperature while the hole concentration for the cross section ${7\times7\:\text{nm}^2}$ is constant in the whole temperature range. Therefore, the size quantum limit may be applied for consideration of the nanowires with the cross section ${7\times7\:\text{nm}^2}$ at the excess hole concentration ${p_{ex}=1\times10^{19}\:\text{cm}^{-3}}$ in the whole temperature range, while it is valid only at low temperatures for a smaller excess hole concentration, for instance, ${p_{ex}=5\times10^{18}\:\text{cm}^{-3}}$.

The Fermi energy for the growth direction [015] is less than that for the [110] direction, nevertheless, the hole concentrations for the both nanowire growth directions coincide at the cross section ${7\times7\:\text{nm}^2}$. In contrast, the hole concentration for the growth direction [015] is somewhat less than that for the other direction at ${p_{ex}=5\times10^{18}\:\text{cm}^{-3}}$, while the Fermi energy is approximately the same for the nanowire cross section ${15\times15\:\text{nm}^2}$. The difference in the Fermi energies and the carrier concentrations associated with different growth directions is owing to the greater confinement effect along the [015] direction. ${E_F}$ for the  ${\bm p\--}$type nanowire significantly changes with temperature compared to that for the intrinsic nanowire. The  ${\bm p\--}$type nanowire hole concentration slightly depends on temperature in comparison with the intrinsic nanowire hole concentration. Both the carrier confinement and the excess holes decrease the temperature dependence of the nanowire carrier concentration; as a result, the intrinsic type of conductivity is suppressed in the ${\bm p\--}$type ${Bi_2Te_3}$ nanowires at high temperatures. Hence, the temperature dependence of the hole (electron) mass can be neglected in the transport calculations. In general, the size quantization decreases the carrier concentration for any type of conductivity of the nanowire. Our calculations show that the confinement effect increases the difference between the Fermi energies as well as between the carrier concentrations corresponding to the growth directions [110] and [015], in contrast to the effect of the excess holes. This leads to a difference in the transport properties of the bismuth telluride nanowires with different growth orientations considered in the next section.  

\section{Thermoelectric-related transport coefficients}
\subsection{Constant-relaxation-time approximation}

The carrier mobility $\mu$ and the relaxation time  ${\tau=\tau_0=m\mu/e}$  are supposed to be constant in the constant-relaxation-time approximation. The electrical conductivity $\sigma$, the Seebeck coefficient $S$, and the electron (hole) thermal conductivity ${\kappa_{e(h)}}$ for a nanowire are defined as \cite{Lin_PRB62, Ashcroft}  
\begin{eqnarray}
\sigma &=& L^{(0)},\\
S &=& -\frac{1}{eT}\frac{L^{(1)}}{L^{(0)}},\\
\kappa_{e(h)} &=& \frac{1}{e^2T}\left[ L^{(2)}-\frac{\left(L^{(1)}\right)^2}{L^{(0)}}\right].
\label{eq:24}
\end{eqnarray}
We use the generalized transport matrix element ${L^{(\alpha)}}$ that presents the sum ${L^{(\alpha)}=\sum\limits_{i}L^{(\alpha)}_i}$  over all six valleys. The transport matrix elements for the \emph{i}th valley have the form \cite{Lin_PRB62, Ashcroft}
\begin{widetext}
\begin{eqnarray}
& L^{(0)}_i  = D_i\sum\limits_{n=1}^{n_{max}}\sum\limits_{l=1}^{l_{max}(n)}\Phi_{-1/2}(\eta^i_{n,l}),\\
&L^{(1)}_i  =  \pm k_B T D_i\sum\limits_{n=1}^{n_{max}} \sum\limits_{l=1}^{l_{max}(n)} \left[  \frac{3}{2} \Phi_{1/2}(\eta^i_{n,l})-\eta^i_{n,l}\Phi_{-1/2}(\eta^i_{n,l})\right],\\
&L^{(2)}_i  =  (k_B T)^2 D_i\sum\limits_{n=1}^{n_{max}} \sum\limits_{l=1}^{l_{max}(n)} \left[   {\frac{15}{4}} \Phi_{3/2}(\eta^i_{n,l})-3\eta^i_{n,l}\Phi_{1/2}(\eta^i_{n,l})+ (\eta^i_{n,l})^2 \Phi_{-1/2}(\eta^i_{n,l}) \right],\\
\label{eq:27} \nonumber
\end{eqnarray}
\end{widetext}
where the sign ${"+"}$ (${"-"}$)corresponds to electrons (holes). The symbol ${D_i}$ denotes the maximal conductivity of a nondegenerate nanowire associated with the \textit{i}th valley, which is expressed as ${D_i=e\mu_y N^{1D}_{i,c(v)}/a_x a_z}$. The electron (hole) thermal conductivity can be written in the form  ${\kappa_{e(h)}=\sigma_{e(h)}LT}$ , where the Lorenz number $L$ for both degenerate one--dimensional and bulk semiconductor compounds is given by ${L^{deg}=(\pi k_B/e)^2/3}$ \cite{Ashcroft}.

For a bipolar nanowire, the transport matrix element includes both electron and hole parts,  ${L^{(\alpha)}=L^{\alpha}_e+L^{\alpha}_h}$. The bipolar Seebeck coefficient ${S_b}$ and the thermal conductivity $\kappa$ are expressed as \cite{Hicks_PRB47_19, Fleurial_JPCS49_2}
 
\begin{eqnarray}
S_b= && \frac{\sigma_e S_e +\sigma_h S_h}{\sigma_e +\sigma_h}, \\
\kappa = && \kappa_L +\kappa_e +\kappa_h + \kappa_{eh} 
\label{eq:29},
\end{eqnarray}
where the last term  ${\kappa_{eh} = T\sigma_e \sigma_h (S_h -S_e)^2/(\sigma_e + \sigma_h)}$ is attributed to the thermal energy carried by the electron-hole pairs generated at the heated end of the sample and moving to the cold end, where they annihilate. The generation and annihilation of the electron-hole pairs are associated with the absorption and emission of the thermal energy, correspondingly \cite{Goltsman}. 

Eq. (29) and (\ref{eq:29}) can be reduced to those for monopolar semiconductor at the appropriate conditions. The first term ${\kappa_L}$ in the right hand side of Eq.~(\ref{eq:29}) is related to the lattice thermal conductivity. At the room temperature, the lattice thermal conductivity measured along the trigonal axis ${\kappa^{[001]}_L}$ and that in the perpendicular direction ${\kappa^{[110]}_L}$ are equal to 0.725 W/mK and 1.45 W/mK, respectively, for the bulk bismuth telluride material \cite{Goldsmid_PPS72, Ainsworth_PPSB69} . It is inversely proportional to temperature, ${\kappa_L=430/T}$, when ${T>50 \: \text K}$ \cite{Walker_PPS76}. For the [015] direction, the lattice thermal conductivity is given by the expression  ${\kappa^{[015]}_L= \cos^2(\theta')\kappa^{[110]}_L + \sin^2(\theta')\kappa^{[001]}_L}$, where ${\theta'}$ denotes the angle between the bisectrix axis and the direction [015] \cite{Fleurial_JPCS49}. This conductivity is equal to 1.25 W/mK at the room temperature. Both theoretical estimations and the measurements of the lattice thermal conductivity of the ${Bi_2Te_3}$ nanowire with diameter 40 nm indicate that the nanowire lattice thermal conductivity is reduced by an order of magnitude from the bulk value \cite{Borca_APL85, Khitun_SM26}. We take into account this fact in our calculations.

While the anisotropy factor of the thermal conductivity is 2, the anisotropy factor of the electrical conductivity is 4 (2.7) for both the ${\bm n (\bm p)\--}$type bismuth telluride compound and the ${Bi_2Te_3}$ film \cite{Fleurial_JPCS49, R_Nature, Ainsworth_PPSB69, Delves_PPS78}. Therefore, the electron (hole) mobility ${\mu^{[001]}_{e(h)}}$ along the trigonal axis is ${300 \: \text{cm}^2/\text{Vs}}$ (${189\: \text{cm}^2/\text{Vs}}$). From the classical definition of the carrier mobility and from Matthiessen's rule for the relaxation time, the electron (hole) mobility is expressed by  ${\mu^{[015]}_{e(h)}= (\cos^2(\theta')/\mu^{[110]}_{e(h)} + \sin^2(\theta')/\mu^{[001]}_{e(h)})^{-1}}$ for the growth direction [015] \cite{Lin_PRB62, Fleurial_JPCS49}. It is equal to ${653\: \text{cm}^2/\text{Vs}}$  (${346\: \text{cm}^2/\text{Vs}}$) for electrons (holes). In the next subsection, we present the temperature dependence of the bismuth telluride nanowire thermoelectric parameters calculated by means of the formulas considered here.

\subsection{Temperature dependence}

\begin{figure*}
\includegraphics{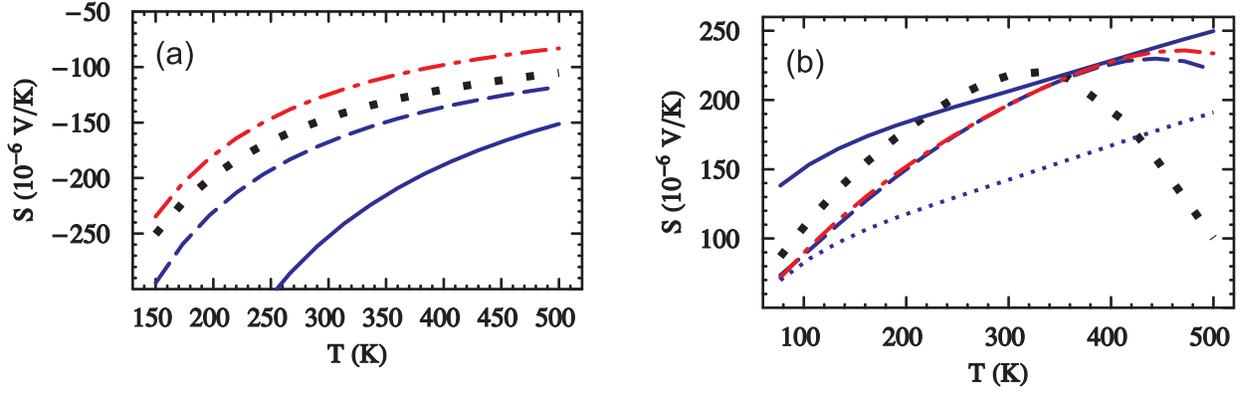}
\caption{\label{fig:5} Temperature dependence of the Seebeck coefficient of (a) the intrinsic and (b) the ${\bm p\--}$type bismuth telluride compound (filled squares) and the square nanowire  with thickness 7 nm (solid line), 15 nm for the growth direction [110] (dashed line) and [015] (dashed-dotted line) at ${p_{ex}=5\times 10^{18}\: \text{cm}^{-3}}$.  The dotted line corresponds to nanowire thickness 7 nm and the growth direction [110] at ${p_{ex}=1\times 10^{19}\: \text{cm}^{-3}}$.}
\end{figure*}

Fig.~\ref{fig:5} shows the temperature dependence of the calculated Seebeck coefficient for the intrinsic and the ${\bm p\--}$type bismuth telluride bulk material and the nanowires with different cross sections, growth directions, and excess hole concentrations. The temperature dependence of the intrinsic nanowire Seebeck coefficient $S$ is monotonic. For the ${\bm p\--}$type nanowire, the temperature dependence of $S$ tends to be monotonic at a small nanowire cross section. In the intrinsic nanowire, the decrease of the cross section leads to an increase of the Seebeck coefficient modulus. The intrinsic Seebeck coefficient is negative because the Fermi energy tends to the conduction subbands owing to the difference between the hole and electron masses. For a high temperature, the calculated value of the intrinsic bulk Seebeck coefficient approaches the measured value  -90 $\mu$V/K, at 700 K \cite{Goltsman}. The calculated maximum value of the Seebeck coefficient ${S=230}$ $\mu$V/K of the bulk ${\bm p\--}$type material shown in Fig.~\ref{fig:5}(b) is less than the measured value ${S=260}$ $\mu$V/K at the room temperature \cite{Goldsmid}. This is because the Seebeck coefficient was measured at the excess hole concentration ${p_{ex}=4\times10^{18}\: \text{cm}^{-3}}$ which is less than the excess hole concentration ${p_{ex}=5\times10^{18}\: \text{cm}^{-3}}$ used in our calculation. In contrast to the ${\bm p\--}$type nanowire, the absolute value of the bipolar Seebeck coefficient for the growth direction [015] is less than that for the direction [110] owing to the carrier mobility difference as it follows from Eq.~(29). In the expression~(24) for the ${\bm p\--}$type Seebeck coefficient, the hole mobility is reduced. An increase of the hole concentration leads to a decrease of the nanowire Seebeck coefficient. For example, $S$ at  ${p_{ex}=5\times10^{18}\: \text{cm}^{-3}}$ is greater by 50 $\mu$V/K than $S$ at ${p_{ex}=1\times10^{19}\: \text{cm}^{-3}}$ in the whole temperature range. In this case, the size quantum limit does not provide a large value of the ${\bm p\--}$type nanowire Seebeck coefficient.
\begin{figure*}
\includegraphics{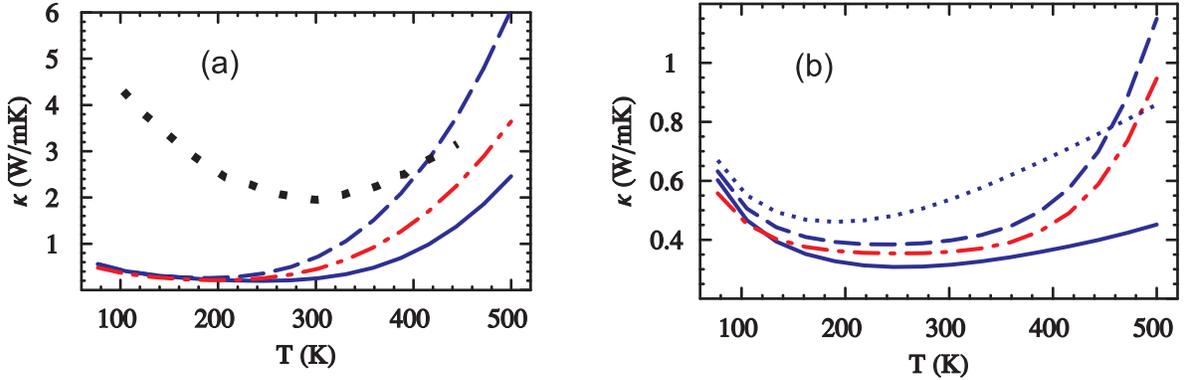}
\caption{\label{fig:6} Temperature dependence of the thermal conductivity in (a) the bulk material (filled squares) \cite{Goltsman}, the intrinsic square nanowire and (b) the ${\bm p\--}$type square nanowire with thickness 7 nm (solid line), 15 nm for the growth direction [110] (dashed line) and [015] (dashed-dotted line). The excess hole concentration is ${p_{ex}=5\times 10^{18}\: \text{cm}^{-3}}$. For the growth direction [110], a dotted line corresponds to the nanowire with thickness 7 nm at  ${p_{ex}=1\times 10^{19}\: \text{cm}^{-3}}$.}
\end{figure*}  

Excess holes suppress the difference between the values of the Seebeck coefficient corresponding to the nanowire growth directions [015] and [110]. In the temperature range from 320 K to 420 K, $S$ for the ${\bm p\--}$type nanowires with different cross sections and the growth directions are similar for the excess hole concentration ${p_{ex}=5\times10^{18}\: \text{cm}^{-3}}$. When temperature ${T>300}$ K, the ${\bm p\--}$type bulk bismuth telluride material tends to be intrinsic; as a result, the bulk Seebeck coefficient strongly decreases \cite{Goltsman}. Both the excess holes and the carrier confinement shift the maximum value of the nanowire Seebeck coefficient towards higher temperatures compared to that for the bulk material, because they suppress the intrinsic type of conductivity. For the excess hole concentration ${p_{ex}=5\times10^{18}\: \text{cm}^{-3}}$, the maximum value of the Seebeck coefficient for the ${\bm p\--}$type square bismuth telluride nanowires with thickness 7 nm and the growth direction [110], as well as with thickness 15 nm corresponding to the growth directions [110] and [015], is equal to 250 $\mu$V/K, 240 $\mu$V/K, and 230 $\mu$V/K, respectively. These values are greater than the respective bulk value of $S$, 230 $\mu$V/K \cite{Goltsman}. At the room temperature, the absolute value of the bipolar nanowire Seebeck coefficient is larger than $S$ for the ${\bm p\--}$type nanowire with a small cross section. The maximum value of $S$ for the intrinsic nanowire is achieved at low temperatures in contrast to that for the ${\bm p\--}$type nanowire.

Fig.~\ref{fig:6} represents the temperature dependence of the total thermal conductivity $\kappa$ of the bismuth telluride compound as well as of the intrinsic and the ${\bm p\--}$type square nanowires. The thermal conductivity increases with temperature because the electron and hole parts of the total thermal conductivity begin to play essential role at high temperatures. The bulk value of $\kappa$ measured along the cleavage plane is less than that for the 7 nm thick nanowire with the corresponding growth orientation at high temperatures. This difference is accounted for by a decreasing temperature dependence of the realistic carrier mobility, which is supposed to be constant in our approximation. The maximum thermal conductivity of the ${\bm p\--}$type nanowire 0.42 W/Km (1.8 W/Km) is less than that of the intrinsic nanowire 2.5 W/Km (6 W/Km) with the same thickness 7 nm (15 nm). This increase is due to the carrier mass temperature dependence in the intrinsic nanowire. As we mentioned above, the phonon part of the nanowire thermal conductivity is less by one order of magnitude than its bulk counterpart.  Therefore, there is a great difference between the values of the thermal conductivity in the bulk and in the nanowire at low temperatures. In contrast to the Seebeck coefficient, the variation in the values of the thermal conductivity increases with temperature for the nanowires with different cross sections and growth orientations. The thermal conductivity of the nanowire with the growth direction [015] is less than that of the nanowire with the direction [110] because of the greater effect of the confinement and another carrier mobility. The nanowires with smaller cross sections have a smaller thermal conductivity. Hence, the carrier confinement causes a decrease of the nanowire thermal conductivity. For the cross section ${7\times7\: \text{nm}^2}$, the thermal conductivity rises with increasing the excess hole concentration. 
\begin{figure*}
\includegraphics{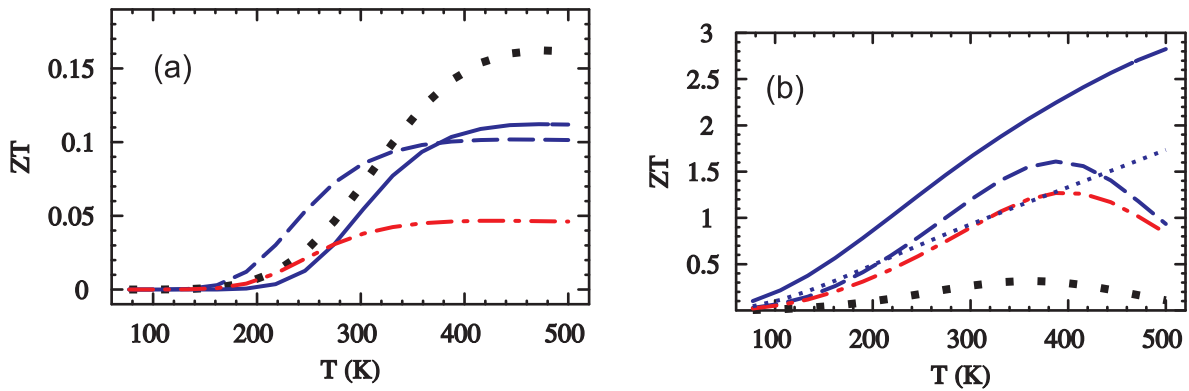}
\caption{\label{fig:7} Temperature dependence of the figure of merit of (a) the bismuth telluride compound (filled squares), the intrinsic square nanowire and (b) the ${\bm p\--}$type square nanowire with thickness 7 nm (solid line) and 15 nm, with the [110] (dashed line) and [015] (dashed-dotted line) growth direction at ${p_{ex}=5\times 10^{18}\: \text{cm}^{-3}}$. Dotted line corresponds to nanowire with thickness 7 nm grown along direction [110], at ${p_{ex}=1\times 10^{19}\: \text{cm}^{-3}}$.}
\end{figure*} 

Fig.~\ref{fig:7} demonstrates the temperature dependence of the figure of merit of the bismuth telluride compound, the intrinsic and the ${\bm p\--}$type square nanowires with different growth directions, the cross sections ${7\times 7\: \text{nm}^2}$ and ${15\times 15\: \text{nm}^2}$, the excess hole concentrations ${p_{ex}=5\times10^{18}\: \text{cm}^{-3}}$ and ${p_{ex}=1\times10^{19}\: \text{cm}^{-3}}$. For the intrinsic nanowire with the growth direction [110], the figure of merit achieves rather small values 0.1 and 0.12 corresponding to the nanowire cross sections ${15\times 15\: \text{nm}^2}$ and ${7\times 7\: \text{nm}^2}$ at temperature 500 K. They are less than the calculated value for the bulk, ${ZT=0.17}$, because the bulk carrier concentration is greater than its nanowire counterpart. At temperature 260 K, the bulk figure of merit 0.03 is less than 0.07 associated with the 15 nm thick intrinsic nanowire due to the smaller nanowire thermal conductivity. When the temperature is below 360 K, ${ZT}$ of the 7 nm-thick nanowire is smaller than that of the 15 nm-thick nanowire, because the difference between the nanowire carrier concentrations is in favor of the latter case. Otherwise, when  ${T>360}$ K, this dependence inverts: a higher value of ${ZT}$ for the 7 nm-thick nanowire is provided by both a larger value of the Seebeck coefficient and a smaller value of the thermal conductivity. Therefore, in the middle of the temperature range under consideration, the carrier concentration mainly influences the figure of merit of the intrinsic nanowire, while at high temperatures, the latter mainly depends on the Seebeck coefficient and the thermal conductivity. Since the carrier concentration and the Seebeck coefficient for the 15 nm-thick intrinsic nanowire grown along the [015] direction are less than those for the bismuth telluride compound and the nanowire with the other growth direction, its figure of merit ${ZT}$ is less than the ${ZT}$ value of the bulk and the nanowire with the growth direction [110]. Hicks et al. \cite{Hicks_PRB47_24,Hicks_PRB47_19} also obtained such a relation from the calculation for the bismuth telluride quantum wells and wires. At the excess hole concentration ${p_{ex}=5\times10^{18}\: \text{cm}^{-3}}$, the calculated bulk figure of merit 0.4 agrees with the value 0.5 reported by Hicks et al. at the room temperature. Otherwise, the figure of merit ${ZT=2}$ for the 7 nm-thick nanowire is considerably greater owing to the smaller thermal conductivity at temperature 300 K. The figure of merit for the ${\bm p\--}$type nanowire increases with decreasing the nanowire cross section. The Seebeck coefficient and the thermal conductivity play the central role in the temperature dependence of the figure of merit for the ${\bm p\--}$type nanowire in the temperature interval under consideration. For the growth direction [110], the maximum value of the figure of merit for the ${\bm p\--}$type nanowire is equal to 1.4, 1.6, and 2.8, correspondingly, at temperatures 310 K, 390 K, and 480 K and the cross sections ${30\times 30\: \text{nm}^2}$, ${15\times 15\: \text{nm}^2}$, and ${7\times 7\: \text{nm}^2}$ (${p_{ex}=5\times10^{18}\: \text{cm}^{-3}}$). At the room temperature, the figure of merit equals 1.2, 1.3, and 1.7, respectively. The maximum value of ${ZT}$ corresponds to the temperature at which the nanowire Seebeck coefficient has its maximum value, while the thermal conductivity begins to increase. This value increases and shifts towards high temperatures when the carrier confinement increases. Our calculations show that the optimal value of the excess hole concentration is about ${p_{ex}=5\times10^{18}\: \text{cm}^{-3}}$. Any deviation from this value leads to a decrease of the figure of merit of the ${\bm p\--}$type nanowire.

\begin{table*}
\caption{\label{tab:3} The calculated thermoelectric parameters for the ${\bm p\--}$type and the intrinsic bismuth telluride nanowires at the room temperature, in comparison with the experimental data on superlattices (SL), nanowires and bulk material.}
\begin{ruledtabular}
\begin{tabular}{lccccc}
 & $S$ & $\kappa$ & $\rho$ & ${ZT}$& Dimensions/ \\
 & ($\mu$V/K)& (W/mK)& (${10^{-5} \Omega\text{m}}$)& & thickness/period \\
\hline
${\bm n}$--type ${Bi_{0.54}Te_{0.46}}$
nanowire \footnote{Reference~\onlinecite{Zhou_APL87}.}
 & -35 & 1 & 1.4 & 0.02 & 81 nm\\
${\bm p\--}$type ${Bi_{0.46}Te_{0.54}}$
nanowire\footnotemark[1] & 260 & - & - & - & 100 nm \\
${\bm n}$--type ${Bi_2Te_3/Bi_2Te_{2.83}Se_{0.17}}$ SL\footnote{Reference~\onlinecite{R_Nature}.}
 & -238 & 0.94 & 1.2 & 1.4 & 6.0 nm\\
${\bm p\--}$type ${Bi_2Te_3/Sb_2Te_3}$ SL\footnotemark[2]\footnote{R. Venkatasubramanian, Phys. Rev. B ${\bm{61}}$, 3091 (2000).}& 165 & 0.65& 0.53& 2.4& 6.0 nm\\
${\bm n}$--type ${Bi_{0.36}Te_{0.64}}$ bulk material\footnote{Reference~\onlinecite{Fleurial_JPCS49}.} 
& -150& 1.8& 1.2& 0.35& 15 mm$\times$15 mm$\times$3 mm\\
${\bm p\--}$type ${Bi_{2}Te_{3}}$ bulk material\footnotemark[4] &
160 & 2.7 & 0.6 & 0.22 & 15 mm$\times$15 mm$\times$3 mm\\ 
Intrinsic ${Bi_{0.37}Te_{0.63}}$ nanowire\footnote{This work.} &
-153 & 0.72 & 11 & 0.085 & 15 nm\\
${\bm p\--}$type ${Bi_{2}Te_{3}}$ nanowire\footnotemark[5] &
143 & 0.53 & 1.2 & 0.94 & 7 nm\\
\end{tabular}
\end{ruledtabular}
\end{table*}
Table \ref{tab:3} contains the calculated thermoelectric parameters and the available experimental data on the thermoelectric properties of the ${\bm p\--}$ and ${\bm n\--}$type bismuth telluride nanowires, superlattices, and bulk materials at the room temperature. The values of the Seebeck coefficient and the thermal conductivity calculated for the ${\bm p\--}$type ${Bi_2Te_3}$ nanowire with thickness 7 nm agree with those measured on the ${\bm p\--}$type ${Bi_2Te_3/Sb_2Te_3}$ superlattice with period 6 nm, while the nanowire electrical resistivity is twice greater than the superlattice resistivity. Since the effect of the one-dimensional confinement on the carrier concentration is less pronounced than the effect of the two-dimensional confinement and our model does not take into account the surface scattering and phonon confinement effect, the calculated value of figure of merit disagrees with that obtained experimentally for the superlattice. It is difficult to make comparison with the experimental data on the nanowires because their diameter is greater than 80 nm when the confinement does not play an essential role.

\section{Conclusions}

The carrier confinement influences the thermoelectric properties of the bismuth telluride nanowires with thickness less than 30 nm. The confinement leads to an increase of the Seebeck coefficient as well as to a decrease of the thermal conductivity and the carrier concentration in both the intrinsic nanowires and the ${\bm p\--}$type nanowires. While the nanowire cross section decreases, the Femri energy, the carrier concentration, the thermal conductivity, and the Seebeck coefficient change more significantly for the growth direction [015] than for the [110] direction because the carrier mass components are smaller for the former direction. In contrast to the ${Bi_2Te_3}$ nanowires, the ${Bi_{0.37}Te_{0.63}}$ nanowires are intrinsic when the temperature ranges from 77 K to 500 K. Therefore, the dependence of the carrier effective mass on temperature should be taken into account in the calculations of transport properties of such nanowires. Six equivalent valleys in the bulk bismuth telluride are split into two-fold and four-fold degenerate valleys for the nanowire growth direction [110] and into 3 doubly degenerate valleys for the direction [015]. Mainly, the excess holes and the carrier confinement have opposite effects on the nanowire thermoelectric parameters. There is an exception, when both the excess holes and the carrier confinement suppress the intrinsic type of electrical conductivity in the ${\bm p\--}$type nanowires at high temperatures owing to a weakening of the carrier concentration dependence on temperature. The size quantum limit can be used if the cross section of the ${\bm p\--}$type nanowire is less than ${8\times 10\: \text{nm}^2}$ (${6\times 7\: \text{nm}^2}$ and ${5\times 5\: \text{nm}^2}$) at the excess hole concentration ${p_{ex}=2\times10^{18}\: \text{cm}^{-3}}$ (${p_{ex}=5\times10^{18}\: \text{cm}^{-3}}$ and ${p_{ex}=1\times10^{19}\: \text{cm}^{-3}}$ correspondingly) in the restricted temperature range. The ${ZT}$ value of the intrinsic nanowire does not exceed the maximal bulk value 0.16, until the thickness of the square nanowire is at least larger than 7 nm. For the growth direction [110], the maximum value of the figure of merit of the ${\bm p\--}$type nanowire is equal to 1.4, 1.6, and 2.8, correspondingly, at temperatures 310 K, 390 K, and 480 K and the cross sections ${30\times 30\: \text{nm}^2}$, ${15\times 15\: \text{nm}^2}$, and ${7\times 7\: \text{nm}^2}$ (${p_{ex}=5\times10^{18}\: \text{cm}^{-3}}$). At the room temperature, the figure of merit equals 1.2, 1.3, and 1.7, respectively. For the ${\bm p\--}$type ${Bi_2Te_3}$ nanowire, the figure of merit ${ZT}$ reaches its maximum value when the excess hole concentration varies in the interval ${p_{ex}=(4 \div 8)\times10^{18}\: \text{cm}^{-3}}$. In general, the carrier confinement increases the figure of merit of the ${\bm p\--}$type bismuth telluride nanowires and shifts its maximum value towards high temperatures. 

\begin{acknowledgments}
We wish to  thank Prof. V. Fomin for his careful reading of an earlier version of our paper and helpful comments. I. Bejenari acknowledges Prof. V. Osipov for his kind invitation to join his research group in the BLTP, JINR from 01/04/07 to 30/06/07.
\end{acknowledgments}

\bibliographystyle{apsrev}

\end{document}